\documentclass[twocolumn,preprintnumbers,amsmath,amssymb]{revtex4-1}
\usepackage{epsfig}
\usepackage{color}
\usepackage{amsmath}

\begin{document}

\title{Seeding Approach to Bubble Nucleation in Superheated Lennard Jones Fluids}

\author{P. Rosales-Pelaez$^1$, M. I. Garcia-Cid$^2$, C. Valeriani$^2$, C. Vega$^1$ and E. Sanz$^1$}
\affiliation{$^1$Departamento de Qu\'{\i}mica F\'{\i}sica,
Facultad de Ciencias Qu\'{\i}micas, Universidad Complutense de Madrid,
28040 Madrid, Spain\\
$^2$ Departamento de Estructura de la Materia, F\'{\i}sica Termica y Electronica,
Facultad de Ciencias  F\'{\i}sicas, Universidad Complutense de Madrid,
28040 Madrid, Spain}

\begin{abstract}
We investigate vapor homogeneous nucleation in a superheated Lennard Jones
liquid with computer simulations.  Special simulation techniques are required to address this 
study since  the nucleation of a critical vapor bubble --one that has
equal chance to grow or shrink-- in a moderately superheated liquid is a rare event. 
We use the Seeding method, that combines Classical Nucleation Theory with computer simulations
of a liquid containing a vapor bubble 
to provide bubble nucleation rates in a wide temperature range. Seeding has been successfully 
applied to investigate the nucleation of crystals in supercooled fluids and here we apply it
for the fist time --to the best of our knowledge-- to the liquid-to-vapor transition.  
We find that Seeding provides nucleation rates that are consistent with independent calculations
not based on the assumptions of Classical Nucleation Theory. Different criteria to determine
the radius of the critical bubble give different rate values. The accuracy
of each criterion depends of the degree of superheating. Moreover, Seeding simulations show
that the surface tension depends on pressure for a given temperature. Therefore, using Classical Nucleation Theory
	\textcolor{black}{with the coexistence surface tension does not provide good estimates of the nucleation rate.} 
\end{abstract}

\maketitle 

\section{Introduction}

The boiling of superheated liquids is important in industrial, technological
and geological processes like explosive boiling
\cite{shusser1999explosive,shusser2000kinetic}, vulcanism
\cite{toramaru1989vesiculation,massol2005effect}, erosion
\cite{brennen2014cavitation} or acoustic cavitation
\cite{suslick1990sonochemistry,suslick1997chemistry}.
In order to control boiling it is important to understand in detail how it works from a molecular perspective. 

The onset of boiling requires the emergence of a critical vapor bubble (cavitation), one sufficiently big to grow.
The experimental study of the formation of such critical bubbles is difficult because they are small ($\sim$ nm) and ephemeral ($\sim$ ns)
and, moreover, their formation is an activated stochastic process, which implies that it is not known a priori where or when 
will a critical bubble appear in the system. 
Molecular simulations have access to such time and length scales and are, therefore, an excellent tool 
to investigate the formation of critical bubbles in molecular detail. 
However, they face the problem of critical bubble formation being an activated process. 
This means that special rare event simulation techniques are required to observe such process in a simulation. 
Thus, Forward Flux Sampling \cite{wang2008homogeneous,meadley2012thermodynamics}, Umbrella Sampling \cite{meadley2012thermodynamics}, or 
brute force Molecular Dynamics simulations with huge systems \cite{tanaka2015simple} 
have been used to study bubble cavitation 
in atomic Lennard Jones fluids, or Umbrella Sampling and Path Sampling have been used to study such phenomenon in molecular 
liquids such as water \cite{menzl2016molecular}. 

All these techniques have been successful in giving accurate values of the
nucleation rate (the number of critical bubbles that form per unit time and
volume) for different types of first order phase transitions 
\cite{searJPCM2007,reviewMichaelides2016,brukhnoJPCM2008,JCP_1998_109_09901,JCP_2005_122_194501,JCP_1992_96_4655,haji-akbariPNAS2015,jiang2018forward}. 
The problem is that they are quite costly from a computational point of view.
Recently, a technique based in combining simulations of a configuration where
the critical nucleus (a bubble in
our case) is inserted in the system from the beginning with Classical Nucleation Theory
\cite{kelton,ZPC_1926_119_277_nolotengo,becker-doring} has proven successful in
providing trends of the nucleation rate in a wide range of orders of magnitude
\cite{seedingvienes}.  This approach, named Seeding, has been successfully
applied to the liquid-to-solid transition
\cite{jacs2013,zaragozaJCP2015,espinosaPRL2016}, but, to the best of our knowledge, it has never been used to study the
liquid-to-vapor one. 

In this paper we apply Seeding for the first time to investigate vapor nucleation. We choose 
to study a Lennard Jones fluid in order to validate 
seeding results with previous calculations of the nucleation rate using more rigorous, but costly, 
techniques \cite{wang2008homogeneous,meadley2012thermodynamics}. 
We find that seeding reasonably predicts the nucleation rate trend and is consistent with previous literature values \cite{wang2008homogeneous,meadley2012thermodynamics} and with
rate values obtained in this work from spontaneous cavitation events at large superheating. 
Depending on the superheating, different criteria to compute the bubble radius
are recommended. At moderate superheating, it works better to identify the bubble radius with the point at which the density is the average between the 
vapor and the liquid ones, whereas at high superheating, a definition based on the equi-molar Gibbs dividing surface gives better results. 
We point out that seeding simulations are needed to obtain reasonable rate estimates because the surface tension that enters 
Classical Nucleation Theory is different from that
at coexistence for a given temperature.

\begin{figure}[h]
\begin{center}
\includegraphics[width=0.85\linewidth]{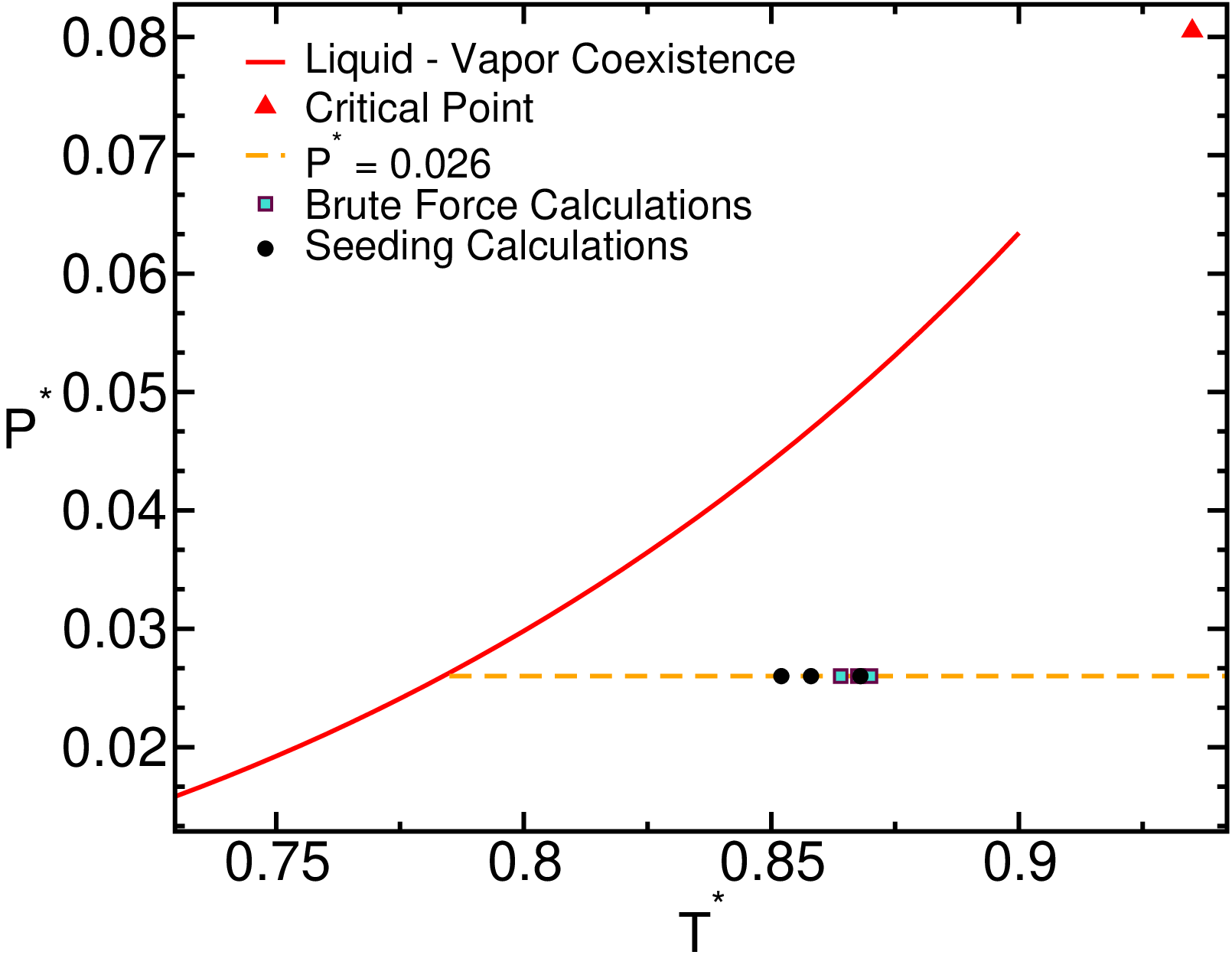}
\caption{Liquid - Vapor phase diagram of the system under study, taken from Ref. 
\cite{wang2008homogeneous}. The isobar for which bubble cavitation has been studied, $P^*=0.026$, 
is indicated with a horizontal dashed line. Black circles
correspond to the conditions where seeding simulations were carried out and turquoise 
squares to brute force simulations of bubble cavitation.}
\label{PT_GR}
\end{center}
\end{figure}

\section{Simulation details}
We carry out 
computer simulations of the truncated and force-shifted Lennard-Jones
(TSF-LJ) potential \cite{wang2008homogeneous}, a model for which 
bubble cavitation has been previously studied \cite{wang2008homogeneous,meadley2012thermodynamics,tanaka2015simple}:

\begin{equation}
U_{TSF-LJ}(r) = U_{LJ}(r) - U_{LJ}(r_{c}) - (r-r_{c})U'_{LJ}(r_{c})
\end{equation}

where $U_{LJ}(r)$ is the 12-6 Lennard-Jones potential and $U'_{LJ}(r)$ is its
first derivative. The interaction potential is truncated and shifted at  $r_{c}=2.5 \sigma$,
being $\sigma$ the particle's diameter.
In what follows, we will use reduced units, expressing all physical variables
in terms of $\sigma$, $\epsilon$, \textcolor{black}{the depth of the Lennard-Jones potential}, and $m$, the mass of the
particles: $T^{*}=k_{B}T \epsilon^{-1}$, $P^{*}=P \cdot \sigma^{3}
\epsilon^{-1}$, $\rho^{*}=\rho \cdot \sigma^{3}$ and $t^{*}=t \left ( \epsilon
m^{-1} \sigma^{-2} \right )^{1/2}$.

All simulations have been performed using the Molecular Dynamics (MD) LAMMPS package \cite{lammps_program},
applying cubic periodic boundary conditions and
integrating the equations of motion with a leap-frog algorithm \cite{leapfrog}
and a timestep of $\Delta t^* = 0.00046$.  Different system sizes have been
used to study different bubble sizes: $N=32000$ for the smallest bubble 
and $N=131072$ for the largest ones (up to \textcolor{black}{a critical radius of 15 $\sigma$}).
To study bubble nucleation, the system has been equilibrated in  an NpT ensemble,
whose temperature and pressure are held constant via a
Nose-Hover thermostat and barostat \cite{JCP_1984_81_00511}
with relaxation
times $\tau_{T}^{*}=0.46$ and $\tau_{P}^{*}=4.6$ respectively.
Our large system sizes allow us to control the pressure with a global barostat despite the fact that 
the system is heterogeneous \cite{marchio2018pressure} and to accommodate the bubble in the simulation box without periodic
boundary effects. \textcolor{black}{In particular, the ratio between the volume of the system and that of the bubble well exceeds
15 (it ranges from $\sim$ 40 to $\sim$ 100), which is roughly the value beyond which the pressure of a heterogeneous system can
be controlled with a global barostat \cite{marchio2018pressure}}.

The pressure-temperature equilibrium phase diagram for the studied model potential is shown in 
Fig. \ref{PT_GR}. The solid line is the coexistence line and the black dots the points
at which we estimated the bubble nucleation rate with seeding. Bubble nucleation was studied
along the $P^{*}=0.026$ isobar, indicated by a horizontal dashed line. Turquoise squares indicate the points
at which we estimated the rate by brute force Molecular Dynamics, which was possible due to the high superheating
of those state points. 

\section{Seeding of cavitation}
\label{seedingofc}

According to 
Classical Nucleation Theory (CNT),
the work needed to form a spherical bubble embedded in 
a metastable liquid at constant pressure and temperature is given by \cite{gibbsCNT1,gibbsCNT2,ZPC_1926_119_277_nolotengo,becker-doring,kelton}:
\begin{equation} 
\Delta G = -V \Delta P + \gamma A_0, 
\end{equation} 
where $V$ is the volume of the bubble, $\Delta P$ the pressure difference 
between the bubble and the surrounding liquid, 
$\gamma$ is the surface tension and $A_0$ is the area of the bubble. 

For spherical bubbles of radius $R$ the expression above becomes:  
\begin{equation} 
\Delta G = -\frac{4}{3} \pi R^{3} \Delta P + \gamma 4 \pi R^{2}.
\end{equation} 
By maximizing this function with respect to $R$ 
the Laplace equation is recovered:
\begin{equation} 
R_{c}=\frac{2 \gamma}{\Delta P}, 
\label{LPeq} 
\end{equation} 
where $R_c$ is the critical bubble radius.
The free energy barrier height corresponding to such radius is:
\begin{equation} 
\Delta G_{c} = \frac{2}{3} \pi R_{c}^{3} \Delta P.
\label{DG_GR} 
\end{equation} 
Thus, knowing the critical bubble radius and the pressure difference, 
we can obtain the free energy barrier height. 

The bubble nucleation rate, or number of critical bubbles that form 
per unit time and volume, can be obtained as:
\begin{equation}
J=A\exp(-\Delta G_c/k_BT),
	\label{eqj}
\end{equation}
where $A$ is the kinetic pre-factor, whose calculation is discussed 
in Section \ref{prefactor}.

In the following sections we explain how to compute $R_c$, $\Delta P$ and $A$ for
a bubble that is critical at $T^*=0.852$.

\section{Bubble radius}
\label{br}

\begin{figure}[h]
\begin{center}
\includegraphics[width=0.7\linewidth]{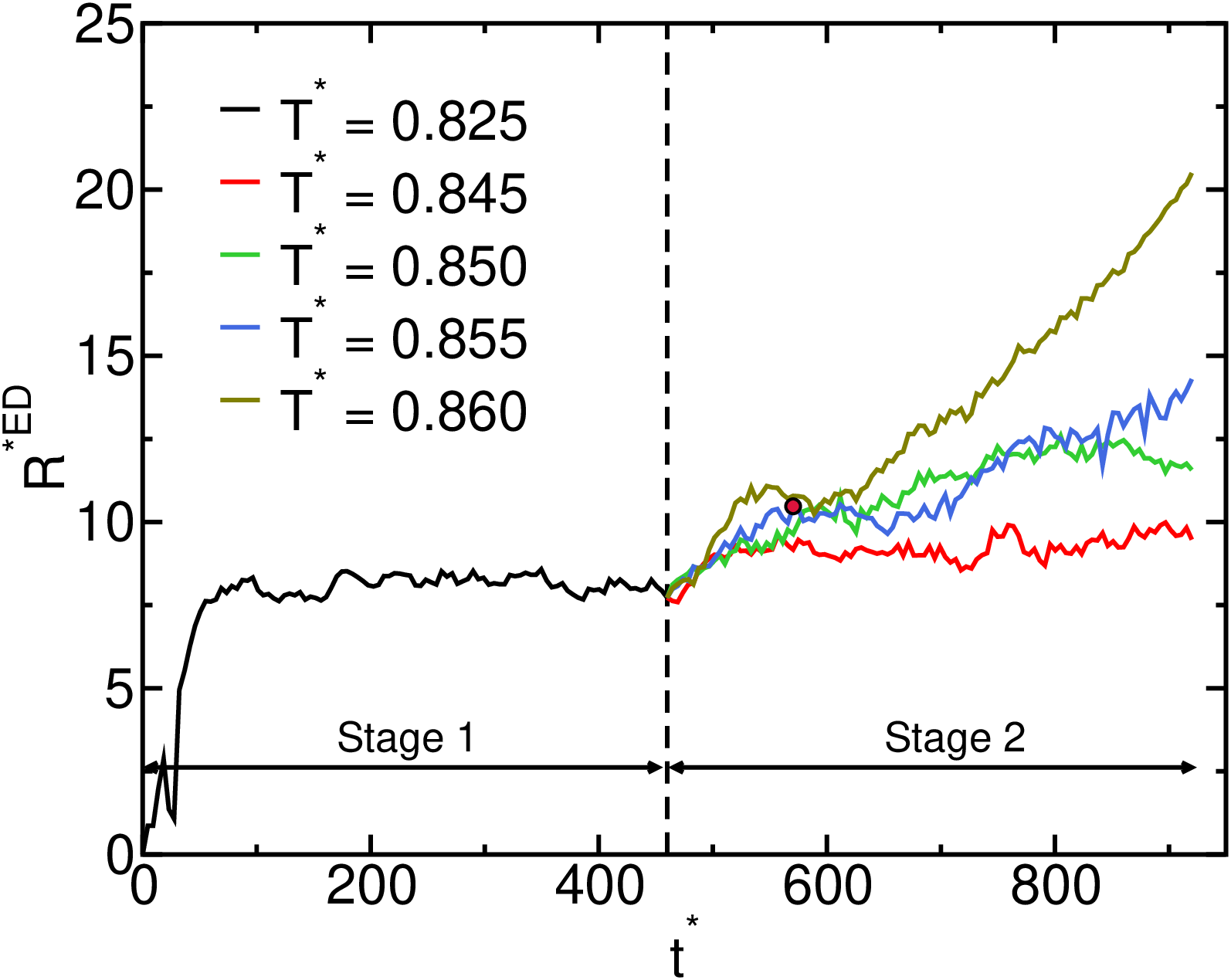}(a)
\includegraphics[width=0.7\linewidth]{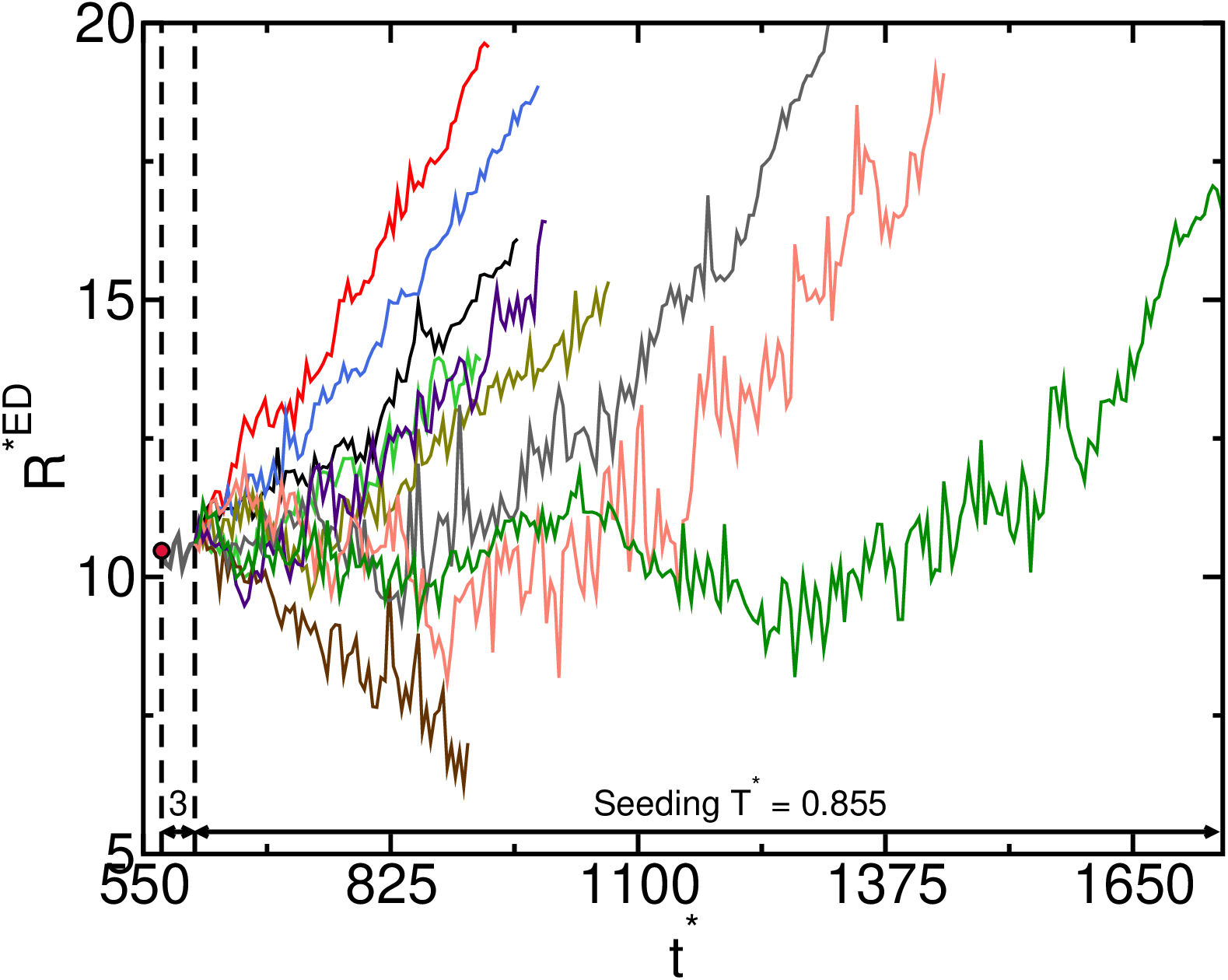}(b)
\caption{
(a) Bubble radius versus time during the growth of a cavity in a fluid at $P^*=0.026$ induced by 
a spherically repulsive potential with radius of 7.5 $\sigma$ and a height of 0.3 $k_BT$. 
Stage 1: The cavity is grown at a temperature at which the bubble does not grow 
(black line, $T^*=0.825$). Stage 2: With the repulsive potential turned on, the final configuration 
of Stage 1 is launched at several temperatures to get a first idea of the temperature that makes the 
grown cavity critical. Trajectories that show immediate growth of the bubble are post-critical
(dark green line, $T^*=0.860$) whereas those for which the bubble does not grow are pre-critical (red line, $T^*=0.845$).
	We take a configuration from a flat region of a trajectory in between these regimes (blue line, $T^*=0.855$) indicated
with a red dot in the figure, as a starting point for the next stage. 
(b) In ``stage 3'' the repulsive potential is turned off and a short trajectory
is launched to ensure that the cavity is filled with an equilibrium vapor density. In the final seeding stage several trajectories (each with a different velocity assignment)
starting from the final configuration of stage 3 are launched at the same temperature, $T^*=0.855$, to get an estimate of the probability of 
	bubble growth at such temperature (0.9 in this case).}  
\label{initgen}
\end{center}
\end{figure}

\begin{figure}[h]
\begin{center}
\includegraphics[width=0.85\linewidth]{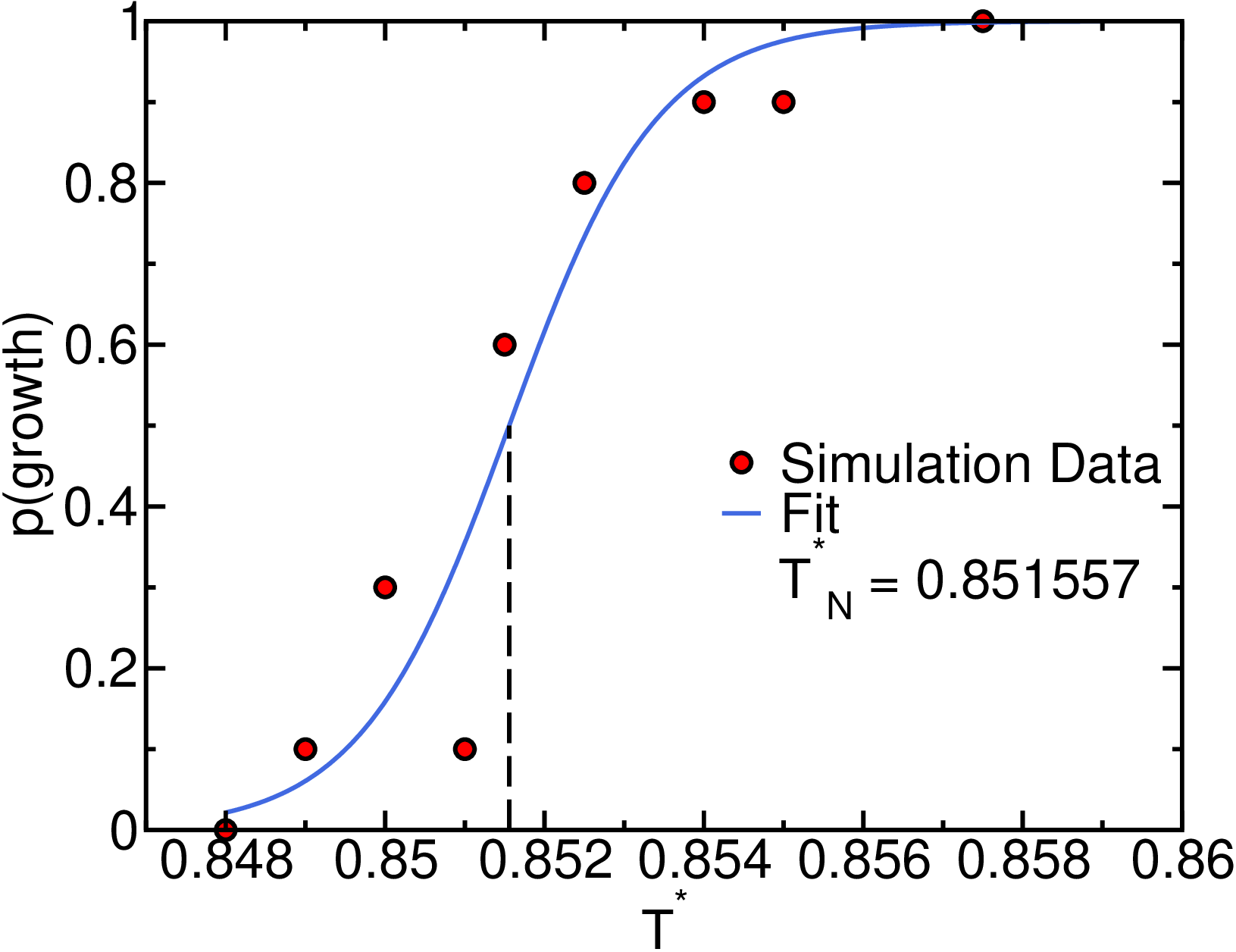}
	\caption{Probability of bubble growth as a function of temperature for a bubble of $R^{ED}=10.5 \sigma$}.
\label{sigmoide}
\end{center}
\end{figure}

To obtain the critical bubble radius we generate a configuration with a bubble
in a superheated fluid from which we  launch MD simulations at different
temperatures to find the temperature at which the bubble is critical.
In the following paragraph we explain how we compute 
the bubble radius for a given configuration. Let us 
continue here describing the procedure to generate a bubble configuration and 
the way we subsequently 
determine the 
conditions for which such bubble is critical. 
In Fig. \ref{initgen} we summarize the steps followed to obtain an initial  
bubble configuration. 
First, we use a spherically symmetric repulsive potential to generate a cavity in the fluid. 
Once the cavity is generated, the repulsive potential is switched off and a short simulation 
is then run to allow for the equilibration of the vapor density inside the generated cavity. 
Finally, we launch several trajectories at different temperatures to obtain the 
probability that the bubble grows as a function of temperature. 
The trajectories for T$^*$=0.855 are shown in Fig. \ref{initgen}(b) and the data for the probability
of bubble growth as a function of temperature are shown in Fig. \ref{sigmoide} (red dots). 
These data are fitted to a sigmoid
function (solid line in Fig. \ref{sigmoide}) and the temperature at which the bubble is critical is that corresponding to  
a probability of one half. This way of determining the temperature at which the bubble is critical 
is similar in spirit to the procedure proposed in Ref. \cite{sigmoide} 
to deal with finite size effects in establishing coexistence points with the 
Direct Coexistence method \cite{laddCPL1977,noyaJCP2008}. 

\begin{figure}[h]
\begin{center}
\includegraphics[width=0.7\linewidth]{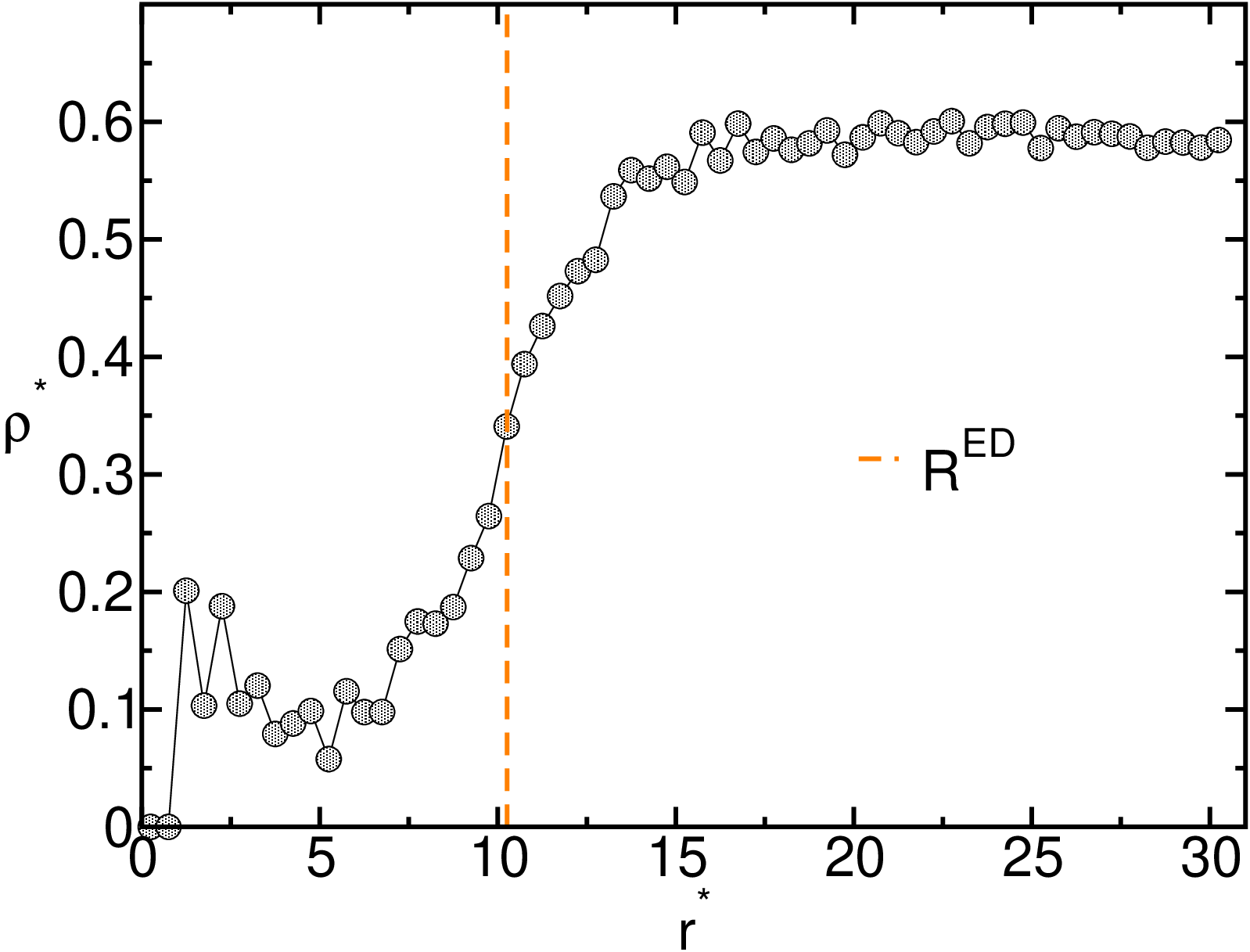}(a)
\includegraphics[width=0.7\linewidth]{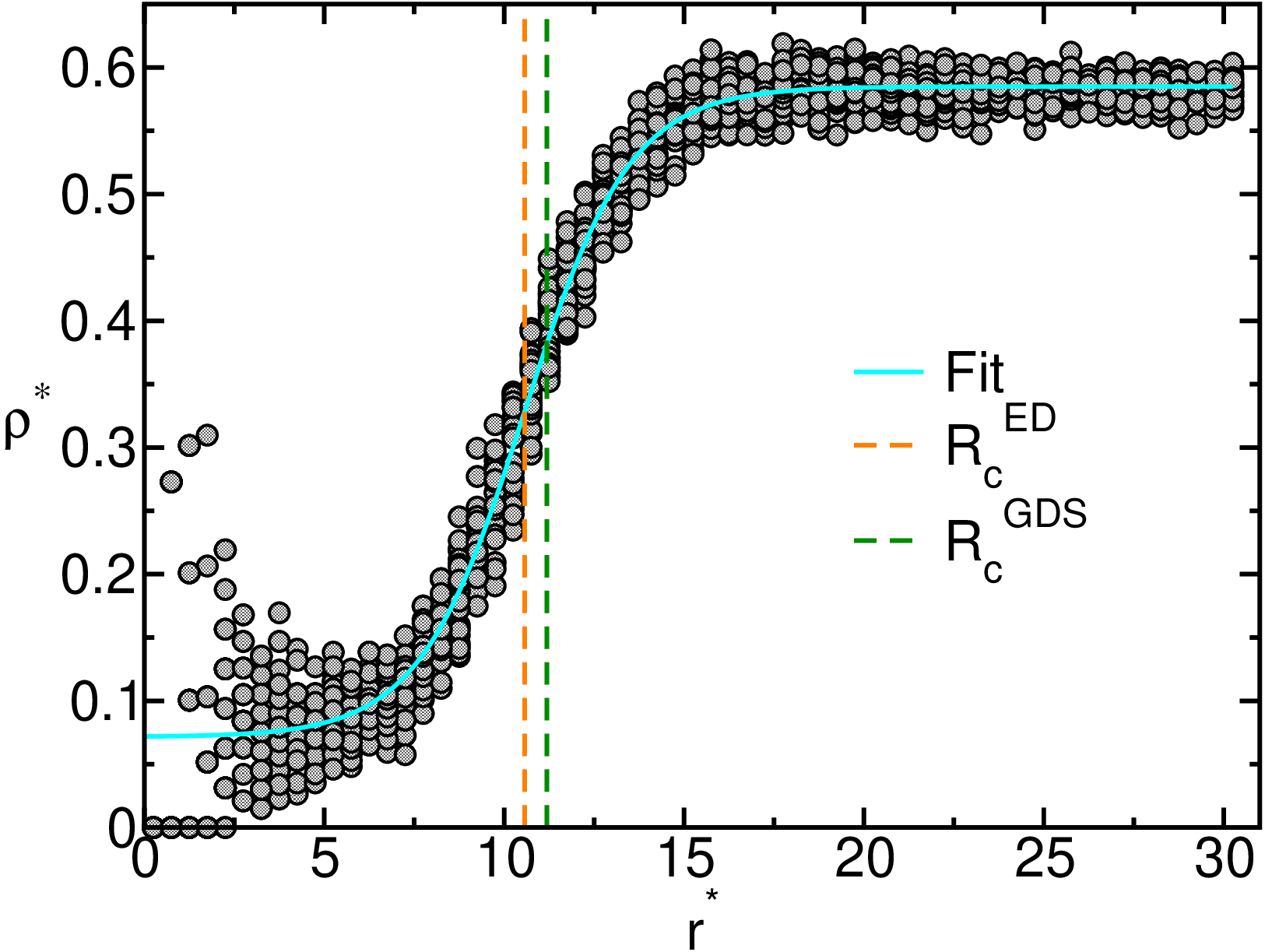}(b)
\caption{(a) Radial density as a function of the distance from the bubble centre
for a given configuration, and (b) for a set of configurations launched from a critical bubble.
Vertical lines indicate the bubble radius according to the ED (dashed orange line) and GDS (dashed green line) criteria.}
\label{bubrad}
\end{center}
\end{figure}

To estimate the bubble's radius in a given configuration we compute a spherically symmetric density profile 
from the bubble's center, $\rho(r)$, as that shown in Fig. \ref{bubrad}(a).
The bubble center coincides with that of the repulsive particle used to generate it. 
We have checked that once the repulsive particle is removed the bubble does not move 
during the time needed to monitor whether the bubble grows or shrinks.  
In Fig. \ref{bubrad}(a) it can be seen that the density is low at short distances, corresponding to the vapor in the interior of the bubble, and high
at long distances, corresponding to the liquid. 
We estimate the radius, $R^{ED}$, as 
the distance at which the density is the average between that in the interior of the bubble, $\rho_v$, and 
that in the liquid, $\rho_l$. We refer to this strategy of finding the bubble radius as the equi-density (ED) method. 
$\rho_l$ and $\rho_v$ are obtained by averaging the density far away from and close to the bubble center respectively.
This is a rough estimate of the radius because $\rho_v$ relies on the first few points of the density profile, which are 
quite noisy (see Fig. \ref{bubrad}(a)). Despite being approximate, this way of obtaining the radius for a single 
configuration enables to quickly determine whether the bubble grows or shrinks. 
In this way we have obtained the bubble radius for all points in Fig. \ref{initgen}, for instance.

The radius for the critical bubble, which is the one that is used to estimate the nucleation rate (see Eqs. \ref{DG_GR} and \ref{eqj}), 
is obtained in a more precise manner. First, we collect density profiles of a few critical bubble configurations, which can be obtained in the short 
period where 
$R^{ED}$ remains constant before either growing or shrinking at the temperature at which the bubble was found to be critical.
Such density profiles for $T^*=0.852$ are shown in Fig. \ref{bubrad}(b). 
$\rho_l$ can be very accurately determined by averaging the density profiles at distances far away from the bubble center. 
$\rho_v$ is trickier to determine \textcolor{black}{by averaging density profiles due to lack of statistics in the interior of the bubble}. 
Instead, $\rho_v$ is obtained by finding the vapor density for which the 
chemical potential is the same in both phases, a condition that is satisfied only for the critical bubble (this way 
to obtain $\rho_v$ is explained in Section \ref{ti}).
Having established $\rho_l$ and $\rho_v$, we assume a sigmoidal function for $\rho(r)$ and fit the density profile data 
to the following function:
\begin{equation}
	\rho(r) = \frac{\rho_{v} + \rho_{l}}{2} + \left ( \frac{\rho_{l} - \rho_{v}}{2} \right ) \cdot \tanh \left [(r - R^{ED}_{c}) / \alpha \right ],
	\label{fit}
\end{equation}
where $\alpha$ is a parameter that is related to the steepness of the interfacial region and $R^{ED}_{c}$
is the radius of the critical bubble for which $\rho(r)=1/2(\rho_v+\rho_l)$, i.e. the equi-density critical radius, which is shown 
by a vertical orange line in Fig. \ref{bubrad}(b).
The fit is shown by a solid line in Fig. \ref{bubrad}(b). 
Note that the horizontal part of the fit close to the center, $\rho_v$, that was imposed in the fitting procedure to 
the value obtained by equating the liquid and the vapor chemical potentials, is consistent with the data
of the computed density profiles. This is a good consistency test proving that we 
are actually dealing with the critical bubble in our simulations. 

Alternatively, we can define the critical bubble radius according 
to the equi-molar Gibbs Dividing Surface (GDS) criterion:
\begin{equation}
	\int_0^{R_{c}^{GDS}} (\rho(r)-\rho_v) r^2 dr = \int_{R_{c}^{GDS}}^{\infty} (\rho_l-\rho(r))r^2dr, 
	\label{gdse}
\end{equation}
where $R_c^{GDS}$ is obtained from the integration limit.
To integrate $\rho(r)$ we use the fit previously obtained (Eq. \ref{fit}). The GDS critical bubble radius is shown by a 
vertical green line in Fig. \ref{bubrad}(b). $R_c^{GDS}$ is larger than $R_c^{ED}$. Obviously, the computed nucleation 
rate will depend on which value of the critical radius is used. Later on, in Section \ref{snr}, we will discuss which radius yields
$J$ values more consistent with independent calculations from the literature.

\section{Pressure difference}
\label{ti}


Obtaining $\Delta P$ is a key step in order to estimate both the interfacial free
energy and the nucleation rate. $\Delta P$ is the difference between the pressure
inside the vapor bubble and the pressure of the metastable liquid around it, i.e.
$\Delta P = P^{V}_{N} - P^{L}_{N}$, where the superscripts $V$ and $L$ refer to 
vapor and liquid respectively, and the subscript $N$ refers to ``nucleation''.  
$P^{L}_{N}$ is simply the pressure imposed in the barostat of our
simulations \textcolor{black}{: $p^*=0.026$. We have checked that this pressure is 
also recovered, with less than 1\% error, by computing the density in the liquid
surrounding the bubble and using the equation of state obtained in independent 
bulk liquid simulations to obtain the pressure.} The pressure inside the bubble can be obtained using the fact
that the chemical potential of the liquid and vapor
phases is the same both at coexistence conditions, $C$, and when a critical bubble is formed at nucleation 
conditions. Then, $\mu^{V}_{N}-\mu^{V}_{C} =
\mu^{L}_{N}-\mu^{L}_{C}$. These chemical potential differences can be obtained by 
integrating the volume per molecule, $V_m$, from the coexistence pressure, $P_C$, to the nucleation one, $P_N$, at the 
temperature at which we found that the inserted bubble is critical:
\begin{equation}
\int_{P_{C}}^{P^{L}_{N}}V^{L}_{m}dP = \int_{P_{C}}^{P^{V}_{N}}V^{V}_{m}dP
\label{TI}
\end{equation}
$V_m$ as a function of pressure can be easily obtained in simulations of the pure liquid and vapor phases. 
The only unknown in the equation above is the upper integration limit of the right-hand-side
integral, $P^{V}_{N}$, which is what we need to obtain $\Delta P$. We obtained for our case study $\Delta P^{*}=0.0163$.
Note that once  $P^{V}_{N}$ is known, it is immediate to obtain $\rho_v$ from the equation of state, which is needed 
to estimate the critical bubble radius according to Eqs. \ref{fit} and \ref{gdse}. 

\begin{figure}[h!]
\begin{center}
\includegraphics[width=0.85\linewidth]{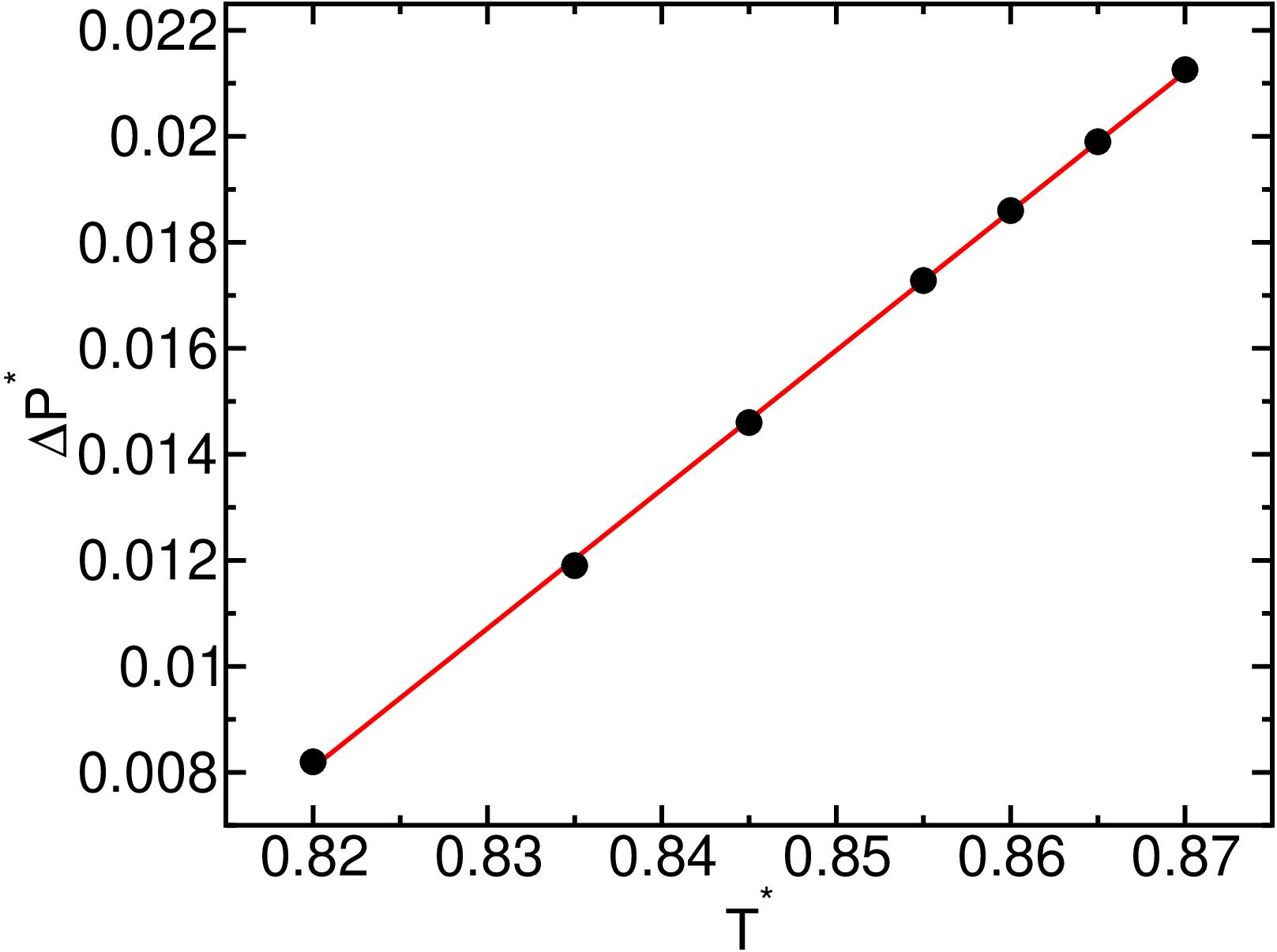}
\caption{Pressure difference between the inside of the bubble and the surrounding liquid, $\Delta P$ as a function of the temperature
for the isobar $P^{*}=0.026$. Black dots correspond to data obtained integrating Eq. \ref{TI} for each temperature and the linear fit for this data,
shown in red, is $\Delta P^{*}= -0.2071 + 0.2624 \cdot T^{*}$.}
\label{DeltaP}
\end{center}
\end{figure}

Moreover, repeating this procedure along the isobar, we observe a linear tendency of $\Delta P$ as a function of the temperature, as
shown in figure \ref{DeltaP}. As discussed later, this linear dependence will be used to fit the nucleation rate as a function of temperature. 

\section{Kinetic pre-factor}
\label{prefactor}

The kinetic prefactor for bubble nucleation is often computed as suggested in
\cite{Blanderino}:
\begin{equation}
A=\rho_l \sqrt{\frac{\Delta P R_{c}}{\pi m}},
\label{J0_1}
\end{equation}
$m$ being the particle mass and $\rho_l$ the number density of the fluid.
Therefore, we can obtain the kinetic pre-factor with the information we have already computed:
 $R_c$ and $\Delta P$.

For our case study, $P^{*}=0.026$ and $T^{*}=0.852$, this equation gives a value for the kinetic pre-factor of $A^{*}=0.139$. 

A \textcolor{black}{second} route to compute the kinetic pre-factor is given in Ref. \cite{menzl2016molecular}:
\begin{equation}
	A = \frac{\sqrt{k_{B}T \Delta P (R_c/2)^{3}}}{\eta} \rho_{l} \rho_{v},
\label{J0_3}
\end{equation}
where $\eta$ is the viscosity of the metastable liquid, $\eta^{*}=0.737$ in our case, 
and $\rho_{l}$ and $\rho_{v}$ are
the number density of liquid and vapor respectively.
The kinetic pre-factor thus obtained is $A^{*}=0.07$.

\begin{figure}[h]
\begin{center}
\includegraphics[width=0.7\linewidth]{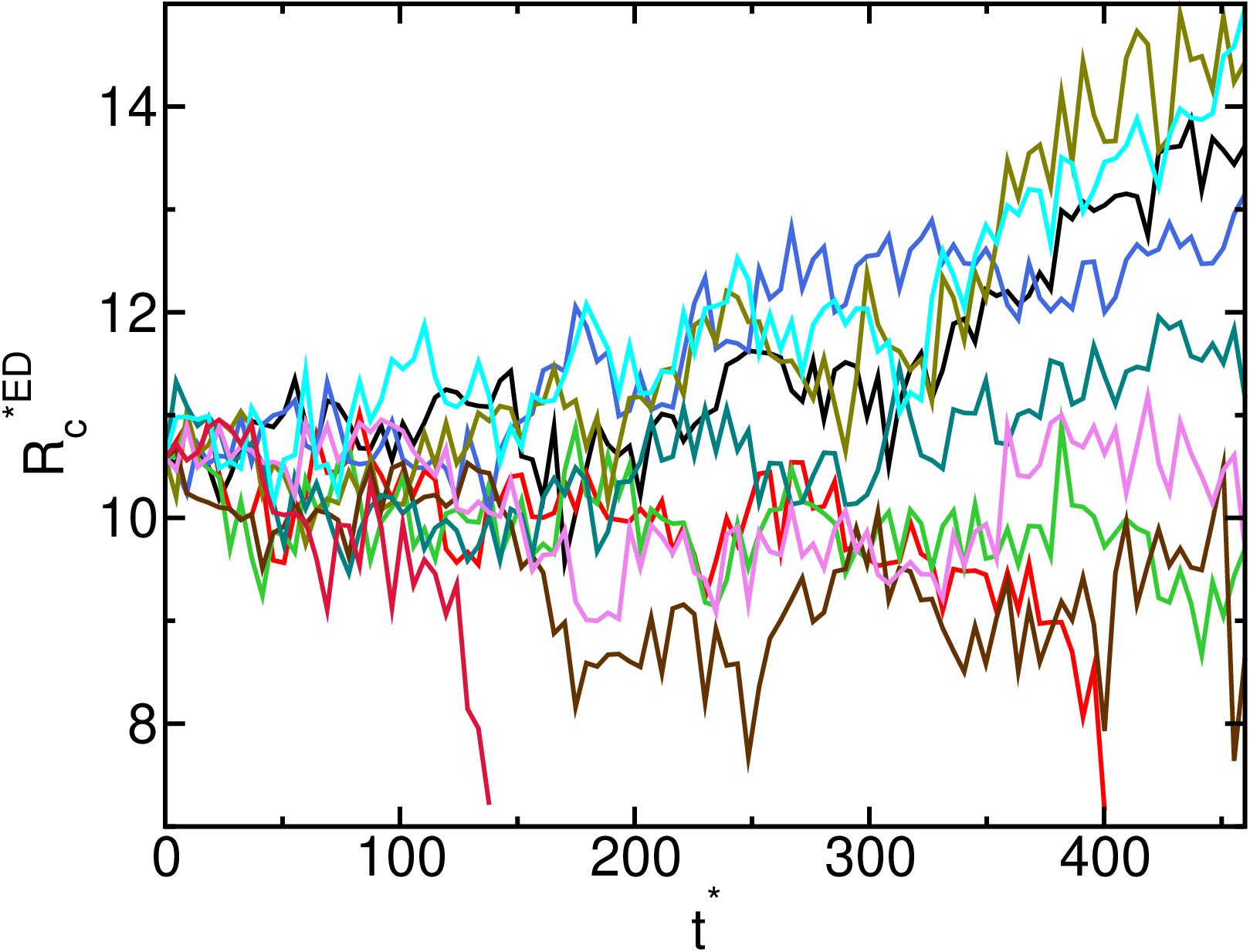}(a)
\includegraphics[width=0.7\linewidth]{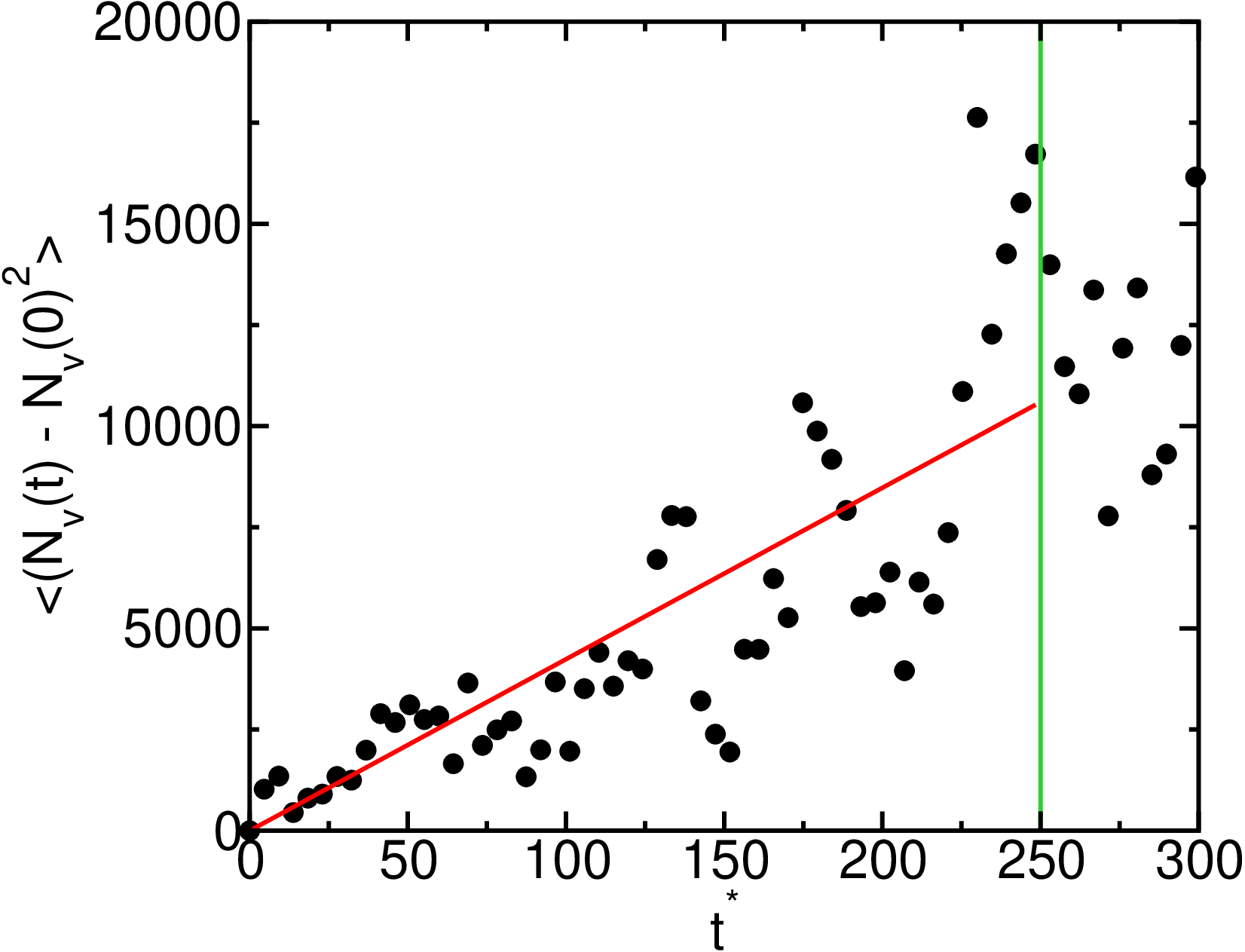}(b)
\caption{
(a): Time evolution of the critical bubble radius for 10  trajectories launched
with different momenta initialization 
starting from a critical bubble at $P^{*}=0.026$ and $T^{*}=0.852$.
(b): Squared deviation of \textcolor{black}{the number of particles in the vapor bubble, $N_v$} with respect to its initial value averaged 
over the 10 trajectories shown in panel (a). The slope of a linear fit for the first
$250 t^{*}$ (red line) gives the attachment rate, $f^{+}$.}
\label{ejemplo_attach}
\end{center}
\end{figure}

Alternatively, we have adapted the expression proposed by Frenkel and Auer for
the kinetic pre-factor for crystal nucleation
\cite{Nature_2001_409_1020,auerJCP2004} to our problem of bubble nucleation.
Auer and Frenkel proposed that the kinetic pre-factor can be obtained via:
\begin{equation}
A=\rho_l \cdot Z \cdot f^{+},
\label{J0_2}
\end{equation}
where $Z$, the Zeldovich factor, is given by \cite{kelton,ZPC_1926_119_277_nolotengo,becker-doring}:
\begin{equation}
Z=\sqrt{\frac{\Delta P}{8 \pi^{2} k_{B}T \rho_{v}^{2} \cdot R_c^{3}}}, 
\end{equation}
$\rho_{v}$ being the number density of vapor particles inside the bubble, obtained from 
the equation of state once we know the pressure inside the bubble from 
the integration of the molecular volumes from coexistence to nucleation conditions, Eq. \ref{TI}. 
$f^+$, is the attachment rate, that can be \textcolor{black}{computed} from the 
expression proposed by Auer and Frenkel \cite{Nature_2001_409_1020,auerJCP2004}:
\begin{equation}
	f^{+} = \frac{\left \langle \left ( N_v(t) - N_v(0) \right )^{2} \right \rangle}{2t}, 
	\label{atteq}
\end{equation}
\textcolor{black}{where $N_v$ is the number of vapor particles in the bubble}. 
To estimate $f^+$ according to this equation we monitor the bubble radius as a function of time, 
$R(t)$ (see \ref{br}), for dozens of molecular dynamics trajectories originated at the
critical bubble. 
Figure \ref{ejemplo_attach}(a)
shows $R(t)$ for our case study. 
The fact that the bubble grows/shrinks in half of the trajectories 
proofs that we are indeed dealing with a critical bubble. 
\textcolor{black}{We can easily convert the bubble radius to $N_v$ assuming a spherical bubble shape
and knowing the vapor density in the critical bubble, $\rho_v$. We obtain the latter from $P^N_V$ (see Eq. \ref{TI}) 
and the vapor equation of state}. 
From such trajectories one can easily obtain the average of the equation above, as shown in 
Fig. \ref{ejemplo_attach}(b). 
The kinetic pre-factor obtained from \textcolor{black}{the slope of Fig. \ref{ejemplo_attach}(b)} and Eq. \ref{J0_2} is $A^{*}=0.09$. 
\textcolor{black}{This approach by Auer and Frenkel to compute the kinetic pre-factor has been used so far for the nucleation of 
crystals. Here, we use it for the first time for vapor nucleation. The good agreement between this approach and the other two previously 
discussed in this section 
suggests that it is reasonable to assume, as Auer and Frenkel did for crystal nucleation, that the growth of the vapor bubble occurs by attachment of 
vapor particles. This view needs to be confirmed with a more careful theoretical treatment, a task that goes beyond the scope 
of our work. Here, we are satisfied with the fact that} the kinetic pre-factors obtained from Eqs. \ref{J0_1}, \ref{J0_3} and \ref{J0_2} 
are of the same order of magnitude (0.139, 0.07 and 0.09 respectively), which is 
quite satisfactory for the calculation of nucleation rates.
We note that these values were computed for $R_c=R_c^{ED}$. If the GDS radius definition had been used instead, the 
values would be slightly different, but within the same order of magnitude.  
\textcolor{black}{It may seem surprising that all three expressions for the rate give similar values in view of the fact that they have different
dependences on the vapor density, $\rho_v$. However, as reported in Table \ref{datos}, the vapor density does not change much from one bubble to another.}

\textcolor{black}{Since all three routes to estimate  
$A$ yield results within the same  order of magnitude we can choose the 
handiest one for the calculation of the nucleation rate (Section \ref{snr}). 
Accordingly, we shall use Eq. \ref{J0_1}, that only depends on $R_c$, $\Delta P$ and
$\rho_l$. The other expressions require costly calculations either of the viscosity (Eq. \ref{J0_3})
or of the attachment rate (Eq. \ref{J0_2}). Moreover, Eq. \ref{J0_1}, combined with the Laplace equation,
gives $A=\rho_l\sqrt{2\gamma/(\pi m)}$, that only depends on $\rho_l$ and $\gamma$. As we show in the following section, $\gamma$ varies
linearly with temperature. Therefore, $A$ can easily be obtained at any temperature in order to fit the 
seeding data for the nucleation rate (see Section \ref{snr}).}

\begin{table*}[t]
\begin{tabular}{|c|c|c|c|c|l|l|l|c|c|c|c|}
\hline
$N$    & $R_{c}^{ED}$ & $R_{c}^{GDS}$ & $T$   & $\Delta P$ & $\gamma_{ED}$ & $\gamma_{GDS}$ & $\rho_{v}$ & $A_{ED}$ & $A_{GDS}$ & $\log(J_{ED})$ & $\log(J_{GDS})$ \\ \hline
32000  & 5.40         & 6.98          & 0.868 & 0.0207     & 0.053       & 0.068        & 0.0786     & 0.105    & 0.119     & -4.2           & -7.9            \\ \hline
131072 & 7.93         & 9.06          & 0.858 & 0.0182     & 0.069       & 0.078        & 0.0728     & 0.125    & 0.133     & -10.0          & -14.5           \\ \hline
131072 & 10.5         & 11.19         & 0.852 & 0.0163     & 0.082       & 0.087        & 0.0715     & 0.139    & 0.144     & -20.1          & -24.2           \\ \hline
\end{tabular}
\caption{Data for the bubble seeding simulations performed in this work at $P^*=0.026$ (in reduced LJ units). 
We report the number or particles in the simulation box, $N$, the bubble radii obtained with two different
criteria, $R_{c}^{ED}$ and $R_{c}^{GDS}$,  the temperature at which the bubble 
is found to be critical, $T$, the pressure difference between the bubble and
	the surrounding liquid, $\Delta P$, \textcolor{black}{the surface tension obtained with the same two criteria, $\gamma_{ED}$ and $\gamma_{GDS}$, the vapor density inside the bubble, $\rho_{v}$,}
the kinetic pre-factor as obtained from Eq. \ref{J0_1} also for two different criteria, $A_{ED}$ and $A_{GDS}$ and the nucleation rates corresponding
	to each criterion to determine the bubble radius, $\log(J_{ED})$ and $\log(J_{GDS})$. All variables are given in reduced units.}
\label{datos}
\end{table*}

\section{Surface tension}

\begin{figure}[h]
\begin{center}
\includegraphics[width=0.85\linewidth]{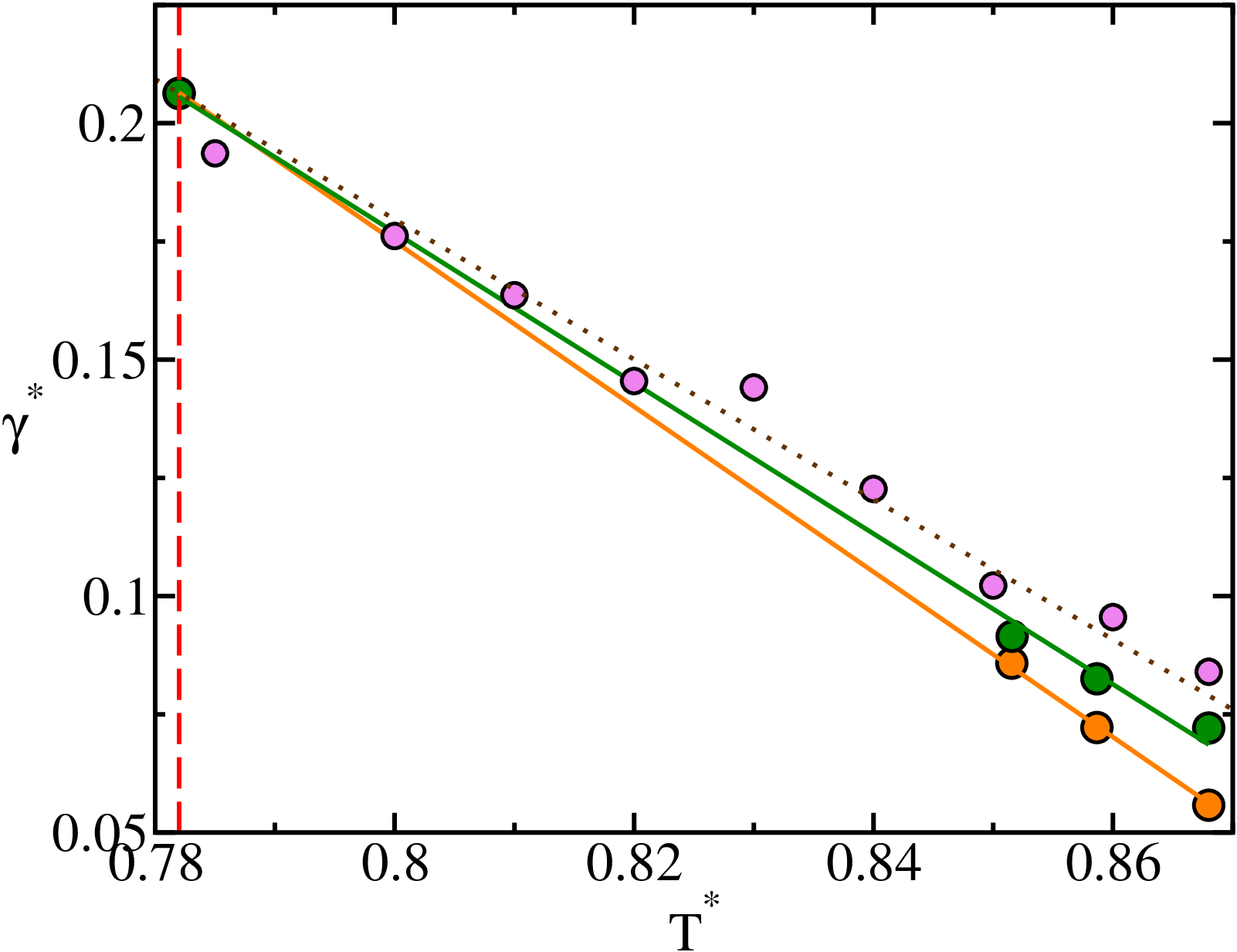}
	\caption{Surface tension as a function of temperature at $P^*=0.026$. Orange and green symbols correspond
	to the ED and GDS definitions of the critical bubble radius respectively. Orange and green lines are linear fits
	to the corresponding data. Pink symbols correspond to data obtained from direct coexistence simulations, $\gamma_{Coex}$, and
	the dotted line corresponds to a linear fit to these data: $\gamma_{Coex}^{*}=1.3644 - 1.4809 \cdot T^{*}$.}
\label{gammafig}
\end{center}
\end{figure}

We can use the Laplace equation (Eq. \ref{LPeq}) to obtain $\gamma$ from $R_c$
and $\Delta P$.  The data thus obtained for each bubble are shown in Fig.
\ref{gammafig}.  There are three sets of data: orange and green, corresponding
to the ED and GDS definition of the critical bubble radius respectively.
$R_c^{ED}$ is lower than $R_c^{GDS}$ and therefore, according to Laplace
equation, yields lower values of $\gamma$. These two sets can be compared to
the third, the pink, corresponding to the surface of tension at the coexistence
pressure for each temperature (reported in Table \ref{gammatab}). This is obtained performing NVT molecular
dynamics simulations of a liquid and a vapor at coexistence,
where the pressure tensor is calculated once the system has equilibrated
following the Irving-Kirkwood expression \cite{walton1983pressure}:
\begin{equation}
\label{gamacoex}
	\gamma = \frac{L_z}{2} (\bar{P}_{z}-\bar{P}_{x,y})
\end{equation}
where $z$ is the direction perpendicular to the liquid-vapor planar interfase and $x$ and $y$ are the tangential directions to the interface.
As shown in Fig. \ref{gammafig}, $\gamma_{Coex}$ 
is higher than that obtained in our seeding simulations. 
The coexistence temperature at the pressure at which
all calculations have been performed in this work, $P^*=0.026$, is indicated by
a vertical dashed line in the figure ($T^*=0.781$).  
A linear
extrapolation of the $\gamma$ seeding data to the coexistence temperature
is consistent with the surface tension at coexistence ($\gamma^{*}_{Coex}=0.21$). 

\textcolor{black}{Our results in Fig. \ref{gammafig} show that $\gamma$ goes down 
as the liquid becomes superheated at constant pressure or undercompressed at constant temperature. 
Accordingly, both routes to obtain a metastable liquid are associated with a positive Tolman length \cite{tolman1949effect,rowlinson2013molecular}, 
$\delta$, in the following expression to describe the variation of $\gamma$ with 
the critical bubble radius:
$\gamma = \gamma_{Coex}\left( 1-\frac{2\delta}{R_c}\right)$, \cite{montero2019interfacial}. It is also clear from Fig. \ref{gammafig} that
the variation of $\gamma$ with pressure at constant temperature is milder than that with temperature at constant pressure. 
Also for the crystal-liquid interface $\gamma$ changes as the system gets away from coexistence \cite{seedingvienes,espinosaPRL2016,montero2019interfacial}.}

\begin{table}[h!]
\begin{tabular}{|c|c|c|c|c|c|c|c|c|c|}
\hline
$T^{*}$      & 0.781 & 0.785 & 0.8   & 0.81  & 0.82  & 0.83  & 0.84  & 0.85  & 0.86  \\ \hline
$\gamma_{Coex}^{*}$ & 0.205 & 0.193 & 0.176 & 0.164 & 0.145 & 0.144 & 0.123 & 0.102 & 0.096 \\ \hline
\end{tabular}
\caption{Surface tension at coexistence as a function of the temperature. Obtained from long coexistence runs and applying Eq. \ref{gamacoex}.}
\label{gammatab}
\end{table}

\section{Nucleation Rate}
\label{snr}
\begin{figure}[h]
\begin{center}
\includegraphics[width=0.85\linewidth]{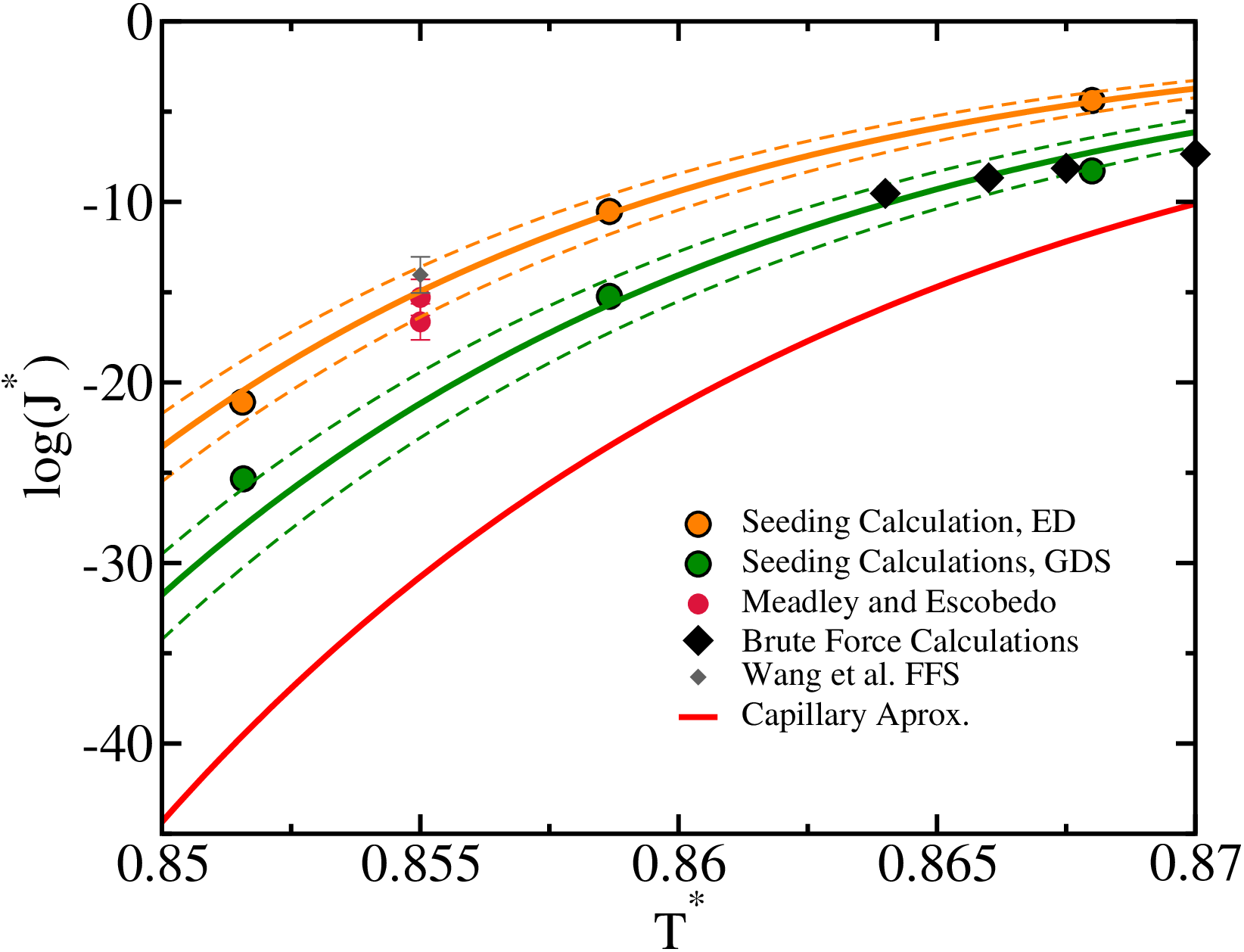}
\caption{Bubble nucleation rate for $P^{*}=0.026$ as a function of temperature. Orange and green dots are seeding results from this work
	with $R_c^{ED}$ and $R_c^{GDS}$ respectively. Solid orange and green lines are 
CNT-inspired fits to the seeding data. Dashed orange and green curves indicate the error bars. 
Red dots \cite{meadley2012thermodynamics} and black diamond \cite{wang2008homogeneous} are literature values. 
Salmon diamonds are results from this work obtained at large superheating by spontaneous cavitation (brute force molecular dynamics). 
The red curve corresponds to the capillary approximation, where 
the coexistence $\gamma$ for each temperature
has been used to estimate the nucleation rate via  Eqs. \ref{LPeq}, \ref{DG_GR}, \ref{eqj} and \ref{J0_1}.}
\label{tasafin}
\end{center}
\end{figure}

With data for the bubble critical radius, the pressure difference between the
fluid and the interior of the bubble and the kinetic pre-factor, we obtain the
nucleation rate as explained in Sec. \ref{seedingofc}.  In Table \ref{datos} we
provide the values of these parameters for the three bubble sizes considered in
this work. \textcolor{black}{As justified in Section \ref{prefactor}, the kinetic pre-factor used to compute 
the rate and reported in Table \ref{datos}} has been computed by means of Eq. \ref{J0_1}.
Two different rates are obtained, one for the equi-density
definition of the bubble radius, $J_{ED}$, and another for the Gibbs Dividing
surface definition, $J_{GDS}$. 
The results for the rates thus obtained are shown with dots
in Fig. \ref{tasafin}. Orange and green symbols correspond to 
$J_{ED}$ and $J_{GDS}$ respectively.
The values for the latter
are lower because $R^{GDS}_c$ is larger than $R^{ED}_c$ (see Fig. \ref{bubrad}(b) and table \ref{datos})
and, according to Eq. \ref{DG_GR}, a larger radius gives raise to a higher
Gibbs free energy barrier and, therefore, to a lower rate.
As critical bubbles become larger (or the superheating decreases) the difference between both radii
decreases. 

We now fit the seeding data of $J(T)$ to a curve inspired on the CNT expressions used to 
compute the nucleation rate. By combining Eqs. \ref{LPeq}, \ref{DG_GR} and \ref{J0_1} the rate can be expressed only
in terms of $\gamma$, $\Delta P$ and $\rho_l$. 
The temperature dependence of the latter is trivially obtained from the equation of state, for that
of $\gamma$ we use the linear fits shown in Fig. \ref{gammafig} and for $\Delta P(T)$ that shown in \ref{DeltaP}. 
Thus, we obtain the solid orange and green curves 
in Fig. \ref{tasafin} that describe the variation of the nucleation rate with temperature. The dashed lines indicate the 
error bars for each $J$ curve, obtained assuming a \textcolor{black}{0.3 $\sigma$} error in the determination of the radius.
\textcolor{black}{The 0.3 $\sigma$ error comes, on the one hand, from the statistical uncertainty in the determination of the radius 
for a given temperature and, on the other hand, from the uncertainty related with the determination of the temperature itself in the seeding simulations. 
Other error sources like $\Delta P$, which is obtained by integrating highly accurate equations of state, or the kinetic pre-factor, 
for which similar values are obtained via three different approaches (see \ref{prefactor}), 
yield a negligible contribution to the rate error as compared to that of the radius}.
The red curve in Fig. \ref{tasafin} is obtained using the coexistence, $\gamma_{Coex}$ for each temperature instead of the $\gamma$ coming
from seeding. We have already shown in Fig. \ref{gammafig} that, for a given temperature, the coexistence $\gamma$ is higher than that obtained from seeding.
Accordingly, the rate curve coming from the coexistence $\gamma$ (in red in Fig. \ref{tasafin}) is lower 
that those coming from seeding (green and orange curves).
This means that the capillary approximation --using $\gamma_{Coex}$ in CNT-- is not valid to obtain nucleation rates. 

The difference between both seeding curves, $J_{GDS}$ and $J_{ED}$, reflects the
ambiguity of the seeding technique with regards to the employed criterion
to determine the size of the critical nucleus \cite{espinosaJCP2016_2}. 
The question is, what is the right value for the radius, $R_c^{ED}$ or
$R_c^{GDS}$?  The answer is neither, because the correct radius is 
the one that provides the correct value for the nucleation rate.
That radius should correspond to the surface of tension, defined as the radius for which Laplace
equation holds \cite{rowlinson2013molecular} (recall from Section \ref{seedingofc} that the Laplace equation is
naturally derived from the CNT formalism behind the seeding technique).
Unfortunately, there is no way to determine the radius of tension from our seeding simulations
since this would require the determination of the free energy of the 
system (i.e there is no rigorous mechanical route to the radius of the surface of tension) \cite{rowlinson2013molecular}.

Therefore, we operationally chose the definition of $R_c$ that gives
values of $J$ closer to those obtained with techniques that do not depend on
CNT. In Fig. \ref{tasafin} we show with black and red symbols the values obtained in
Refs. \cite{meadley2012thermodynamics,wang2008homogeneous} for a given state
point, $T^*=0.855$ and $P^{*}=0.026$, with rare event techniques like Forward Flux Sampling \cite{PRL_2005_94_018104} 
or Umbrella Sampling \cite{JCP_1992_96_04655}. 
These literature values are in better agreement with the
seeding fit obtained with the ED definition of $R_c$. Therefore,
it seems from this comparison that 
$R_{c}^{ED}$ gives the correct $J$ (i. e. that the surface of tension is closer to the ED than to the GDS). 

\begin{figure}[h]
\begin{center}
\includegraphics[width=0.85\linewidth]{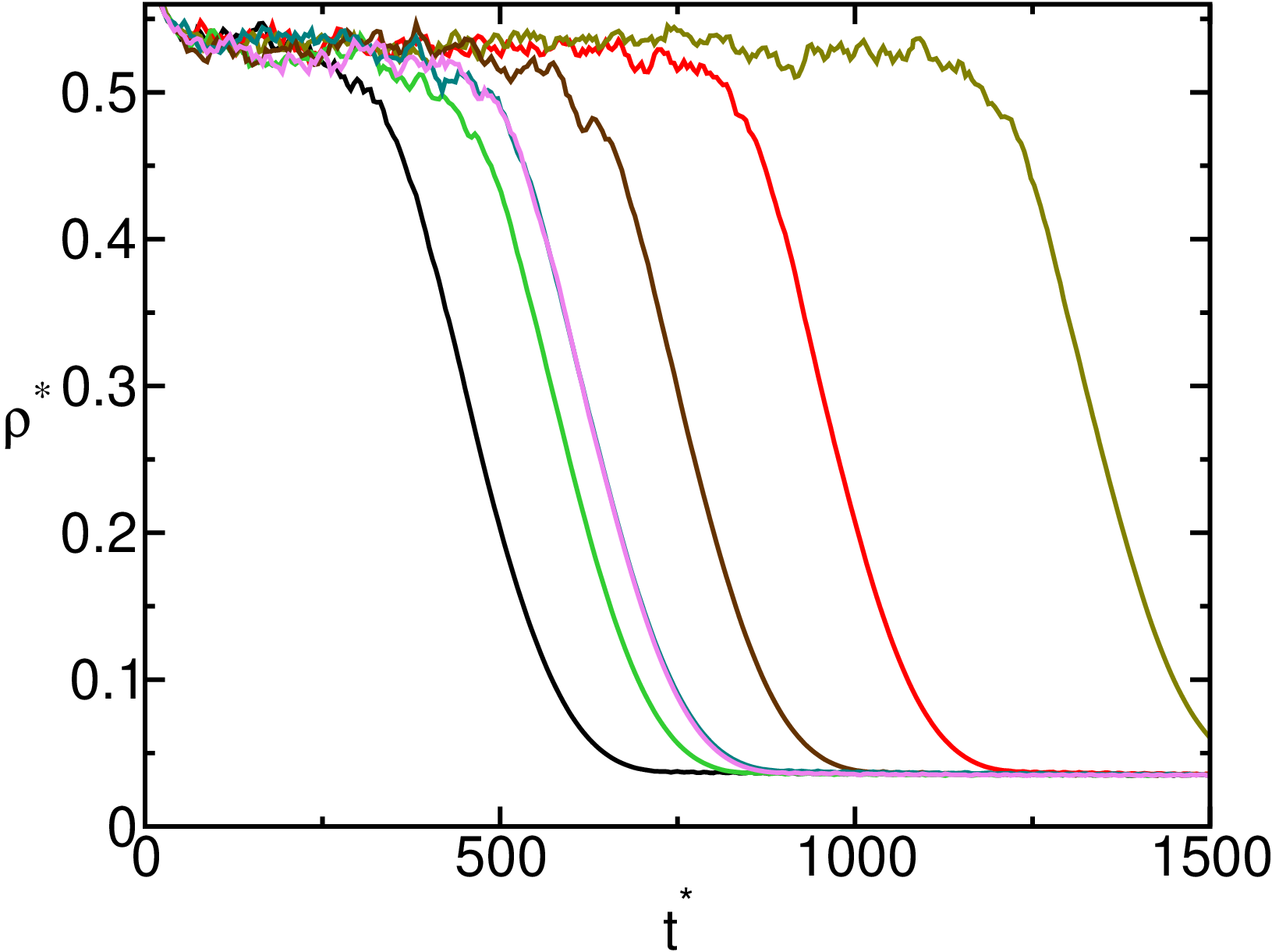}
	\caption{Total density of the system versus time for several trajectories starting from an equilibrated fluid at $T^*=0.87$ and $P^{*}=0.026$.}
\label{espofig}
\end{center}
\end{figure}

However, this conclusion is contradictory with that drawn from comparing seeding data 
with those obtained from spontaneous cavitation at large superheating. 
In such conditions the formation of a critical bubble is spontaneous and the rate can be computed as \cite{filion:244115}:
\begin{equation}
	J=\frac{1}{<t>V}
\end{equation}
where $<t>$ is the average time required to observe cavitation, which is detected by a sudden density drop. 
In Fig. \ref{espofig} we show the density as a function of time for several trajectories at $T^*=0.87$ and $P^{*}=0.026$. 
The stochastic nature of the formation of a critical bubble is evident from the dispersion of cavitation times in the figure. 
In this way we have computed the nucleation rate for several state points at large superheating. The results are shown with 
salmon diamonds in Fig. \ref{tasafin} and are reported in table \ref{brutetable}.

\begin{table}[h!]
\begin{tabular}{|c|c|c|c|c|}
\hline
$T^{*}$      & 0.8640 & 0.8660 & 0.8675 & 0.8700 \\ \hline
$log(J^{*})$ & -9.53  & -8.66  & -8.14  & -7.35  \\ \hline
\end{tabular}
\caption{Brute Force calculations of the nucleation rate at different temperatures and $P^{*}=0.026$.}
\label{brutetable}
\end{table}

Somewhat frustratingly, the seeding curve that better agrees with the spontaneous cavitation data is that coming from the GDS
definition of the critical bubble radius.
 This means that the 
surface of tension, that would provide 
the true value for the nucleation rate, is neither provided by the ED nor the GDS.
Therefore, there is no single seeding curve able to describe the nucleation rate in 
the whole temperature range. The ED curve works better for moderate superheating, whereas the GDS one is more accurate at large
superheating.
\textcolor{black}{It is worth recalling here that seeding 
is expected to work better for large bubbles (small superheating) given that it relies on the CNT assumption
that the bubble has thermodynamic entity.
In this respect, the radius that reproduces the data at low superheating, $R^{ED}$, is the most reliable 
for making seeding estimates of the nucleation rate. $R^{GDS}$ is not so useful then.  
After all, at large superheating, where bubble cavitation is spontaneous, there is no need for seeding predictions. 
}.
In any case it is worth noting that seeding is an approximate technique whose strength is its ability to provide 
values of the rate for a large range of orders of magnitude, at the expense of accuracy. In this respect, \textcolor{black}{$R^{ED}$
yields reasonably good results in the whole temperature range. In fact, the difference between seeding and spontaneous nucleation 
data is of the same order of magnitude as that previously found in seeding studies of crystallization \cite{seedingvienes}.} 

\section{Summary and conclusions}

We investigate bubble cavitation in a superheated Lennard Jones fluid with computer simulations using the Seeding method.
We prepare configurations of a fluid at $P^{*}=0.026$ containing a vapor bubble, and launch several Molecular Dynamics 
trajectories 
at different temperatures to find that at which the generated bubble is critical 
(i. e. it has a growth chance of 1/2). 
We calculate the bubble radius with the aid of radial density profiles centred at the bubble and
the pressure inside the bubbles with thermodynamic integration assuming that the chemical potential of the 
vapor and the liquid is the same for the critical bubble. 
With this information we estimate the free energy required to form the critical bubble 
via Classical Nucleation Theory. The exponential of such barrier combined with a kinetic pre-factor, that 
we obtain with three different approaches consistent among themselves, gives us an estimate of the bubble nucleation rate. 
The seeding nucleation rate is in fairly good agreement with independent calculations using techniques that
do not depend on Classical Nucleation Theory. We use two different ways of identifying the bubble radius. 
One based on the point at which the density is the average between that of the vapor and of the liquid, and another
based on the equi-molar Gibbs dividing surface. 
\textcolor{black}{We conclude that the density based radius definition is more reliable for making seeding predictions
because it works better in the regime where CNT assumptions, on which seeding relies, are more justifiable}.
Our simulations show that 
using the surface tension at coexistence for a given 
temperature to estimate nucleation rates does not give good results because the surface tension 
varies with pressure for a given temperature (or with temperature for a given pressure). Therefore, seeding simulations
are needed to get reasonable nucleation rates in combination with Classical Nucleation Theory. 
Finally, we note that the seeding technique is quite powerful: working with three different bubble sizes we were able to estimate
nucleation rates in a range of tens of orders of magnitude. 

\section{Acknowledgments}
This work was funded by grant FIS2016/78117-P of the MEC.
C. Valeriani thanks financial support from FIS2016-78847-P of the MEC. 
The authors acknowledge the computer resources and technical assistance provided by the RES and L. G. MacDowell for useful discussions.

\clearpage


\begin{thebibliography}{10}

\bibitem{shusser1999explosive}
M.~Shusser and D.~Weihs, ``Explosive boiling of a liquid droplet,'' {\em
  International journal of multiphase flow}, vol.~25, no.~8, pp.~1561--1573,
  1999.

\bibitem{shusser2000kinetic}
M.~Shusser, T.~Ytrehus, and D.~Weihs, ``Kinetic theory analysis of explosive
  boiling of a liquid droplet,'' {\em Fluid Dynamics Research}, vol.~27, no.~6,
  p.~353, 2000.

\bibitem{toramaru1989vesiculation}
A.~Toramaru, ``Vesiculation process and bubble size distributions in ascending
  magmas with constant velocities,'' {\em Journal of Geophysical Research:
  Solid Earth}, vol.~94, no.~B12, pp.~17523--17542, 1989.

\bibitem{massol2005effect}
H.~Massol and T.~Koyaguchi, ``The effect of magma flow on nucleation of gas
  bubbles in a volcanic conduit,'' {\em Journal of volcanology and geothermal
  research}, vol.~143, no.~1-3, pp.~69--88, 2005.

\bibitem{brennen2014cavitation}
C.~E. Brennen, {\em Cavitation and bubble dynamics}.
\newblock Cambridge University Press, 2014.

\bibitem{suslick1990sonochemistry}
K.~S. Suslick, ``Sonochemistry,'' {\em science}, vol.~247, no.~4949,
  pp.~1439--1445, 1990.

\bibitem{suslick1997chemistry}
K.~S. Suslick, M.~M. Mdleleni, and J.~T. Ries, ``Chemistry induced by
  hydrodynamic cavitation,'' {\em Journal of the American Chemical Society},
  vol.~119, no.~39, pp.~9303--9304, 1997.

\bibitem{wang2008homogeneous}
Z.-J. Wang, C.~Valeriani, and D.~Frenkel, ``Homogeneous bubble nucleation
  driven by local hot spots: A molecular dynamics study,'' {\em The Journal of
  Physical Chemistry B}, vol.~113, no.~12, pp.~3776--3784, 2008.

\bibitem{meadley2012thermodynamics}
S.~L. Meadley and F.~A. Escobedo, ``Thermodynamics and kinetics of bubble
  nucleation: Simulation methodology,'' {\em The Journal of chemical physics},
  vol.~137, no.~7, p.~074109, 2012.

\bibitem{tanaka2015simple}
K.~K. Tanaka, H.~Tanaka, R.~Angelil, and J.~Diemand, ``Simple improvements to
  classical bubble nucleation models,'' {\em Physical review E}, vol.~92,
  no.~2, p.~022401, 2015.

\bibitem{menzl2016molecular}
G.~Menzl, M.~A. Gonzalez, P.~Geiger, F.~Caupin, J.~L. Abascal, C.~Valeriani,
  and C.~Dellago, ``Molecular mechanism for cavitation in water under
  tension,'' {\em Proceedings of the National Academy of Sciences}, vol.~113,
  no.~48, pp.~13582--13587, 2016.

\bibitem{searJPCM2007}
R.~P. Sear, ``Nucleation: theory and applications to protein solutions and
  colloidal suspensions,'' {\em Journal of Physics: Condensed Matter}, vol.~19,
  no.~3, p.~033101, 2007.

\bibitem{reviewMichaelides2016}
G.~C. Sosso, J.~Chen, S.~J. Cox, M.~Fitzner, P.~Pedevilla, A.~Zen, and
  A.~Michaelides, ``Crystal nucleation in liquids: Open questions and future
  challenges in molecular dynamics simulations,'' {\em Chemical Reviews},
  vol.~116, no.~12, pp.~7078--7116, 2016.

\bibitem{brukhnoJPCM2008}
A.~V. Brukhno, J.~Anwar, R.~Davidchack, and R.~Handel, ``Challenges in
  molecular simulation of homogeneous ice nucleation,'' {\em Journal of
  Physics: Condensed Matter}, vol.~20, no.~49, p.~494243, 2008.

\bibitem{JCP_1998_109_09901}
P.~R. ten Wolde and D.~Frenkel, ``Computer simulation study of gas-liquid
  nucleation in a {L}ennard-{J}ones system,'' {\em J. Chem. Phys.}, vol.~109,
  p.~9901, 1998.

\bibitem{JCP_2005_122_194501}
C.~Valeriani, E.~Sanz, and D.~Frenkel, ``Rate of homogeneous crystal nucleation
  in molten {NaCl},'' {\em J. Chem. Phys.}, vol.~122, p.~194501, 2005.

\bibitem{JCP_1992_96_4655}
J.~S. van Duijneveld and D.~Frenkel, ``Computer simulation study of free energy
  barriers in crystal nucleation,'' {\em J. Chem. Phys.}, vol.~96, p.~4655,
  1992.

\bibitem{haji-akbariPNAS2015}
A.~Haji-Akbari and P.~G. Debenedetti, ``Direct calculation of ice homogeneous
  nucleation rate for a molecular model of water,'' {\em Proceedings of the
  National Academy of Sciences}, vol.~112, no.~34, pp.~10582--10588, 2015.

\bibitem{jiang2018forward}
H.~Jiang, A.~Haji-Akbari, P.~G. Debenedetti, and A.~Z. Panagiotopoulos,
  ``Forward flux sampling calculation of homogeneous nucleation rates from
  aqueous nacl solutions,'' {\em The Journal of Chemical Physics}, vol.~148,
  no.~4, p.~044505, 2018.

\bibitem{kelton}
K.~F. Kelton, {\em Crystal Nucleation in Liquids and Glasses}.
\newblock Boston: Academic, 1991.

\bibitem{ZPC_1926_119_277_nolotengo}
M.~Volmer and A.~Weber, ``Keimbildung in ubersattigten gebilden,'' {\em Z.
  Phys. Chem.}, vol.~119, p.~277, 1926.

\bibitem{becker-doring}
R.~Becker and W.~Doring, ``Kinetische behandlung der keimbildung in
  ubersattigten dampfen,'' {\em Ann. Phys.}, vol.~416, pp.~719--752, 1935.

\bibitem{seedingvienes}
J.~R. Espinosa, C.~Vega, C.~Valeriani, and E.~Sanz, ``Seeding approach to
  crystal nucleation,'' {\em J. Chem. Phys.}, vol.~144, p.~034501, 2016.

\bibitem{jacs2013}
E.~Sanz, C.~Vega, J.~R. Espinosa, R.~Caballero-Bernal, J.~L.~F. Abascal, and
  C.~Valeriani, ``Homogeneous ice nucleation at moderate supercooling from
  molecular simulation,'' {\em Journal of the American Chemical Society},
  vol.~135, no.~40, pp.~15008--15017, 2013.

\bibitem{zaragozaJCP2015}
A.~Zaragoza, M.~M. Conde, J.~R. Espinosa, C.~Valeriani, C.~Vega, and E.~Sanz,
  ``Competition between ices ih and ic in homogeneous water freezing,'' {\em
  The Journal of Chemical Physics}, vol.~143, no.~13, p.~134504, 2015.

\bibitem{espinosaPRL2016}
J.~R. Espinosa, A.~Zaragoza, P.~Rosales-Pelaez, C.~Navarro, C.~Valeriani,
  C.~Vega, and E.~Sanz, ``Interfacial free energy as the key to the
  pressure-induced deceleration of ice nucleation,'' {\em Phys. Rev. Lett.},
  vol.~117, p.~135702, 2016.

\bibitem{lammps_program}
S.~Plimpton {\em J. Comput. Phys.}, vol.~117, p.~1, 1995.

\bibitem{leapfrog}
R.~W. Hockney, S.~P. Goel, and J.~Eastwood, ``Quiet high resolution computer
  models of a plasma,'' {\em J. Comp. Phys.}, vol.~14, pp.~148--158, 1974.

\bibitem{JCP_1984_81_00511}
S.~Nosé, ``A unified formulation of the constant temperature molecular
  dynamics methods,'' {\em J. Chem. Phys.}, vol.~81, p.~511, 1984.

\bibitem{marchio2018pressure}
S.~Marchio, S.~Meloni, A.~Giacomello, C.~Valeriani, and C.~Casciola, ``Pressure
  control in interfacial systems: Atomistic simulations of vapor nucleation,''
  {\em The Journal of chemical physics}, vol.~148, no.~6, p.~064706, 2018.

\bibitem{gibbsCNT1}
J.~W. Gibbs, ``On the equilibrium of heterogeneous substances,'' {\em Trans.
  Connect. Acad. Sci.}, vol.~3, pp.~108--248, 1876.

\bibitem{gibbsCNT2}
J.~W. Gibbs, ``On the equilibrium of heterogeneous substances,'' {\em Trans.
  Connect. Acad. Sci.}, vol.~16, pp.~343--524, 1878.

\bibitem{sigmoide}
J.~R. Espinosa, E.~Sanz, C.~Valeriani, and C.~Vega, ``On fluid-solid direct
  coexistence simulations: The pseudo-hard sphere model,'' {\em The Journal of
  Chemical Physics}, vol.~139, no.~14, p.~144502, 2013.

\bibitem{laddCPL1977}
A.~Ladd and L.~Woodcock, ``Triple-point coexistence properties of the
  lennard-jones system,'' {\em Chemical Physics Letters}, vol.~51, no.~1,
  pp.~155 -- 159, 1977.

\bibitem{noyaJCP2008}
E.~G. Noya, C.~Vega, and E.~de~Miguel, ``Determination of the melting point of
  hard spheres from direct coexistence simulation methods,'' {\em The Journal
  of Chemical Physics}, vol.~128, no.~15, p.~154507, 2008.

\bibitem{Blanderino}
{Blander, Milton, and Joseph L. Katz. }, ``{Bubble nucleation in liquids.},''
  {\em AIChE Journal}, vol.~21.5, pp.~33--848, 1975.

\bibitem{Nature_2001_409_1020}
S.~Auer and D.~Frenkel, ``Prediction of absolute crystal-nucleation rate in
  hard-sphere colloids,'' {\em Nature}, vol.~409, p.~1020, 2001.

\bibitem{auerJCP2004}
S.~Auer and D.~Frenkel, ``Numerical prediction of absolute crystallization
  rates in hard-sphere colloids,'' {\em J. Phys.:Condens. Matter}, vol.~120,
  p.~3015, 2004.

\bibitem{walton1983pressure}
J.~Walton, D.~Tildesley, J.~Rowlinson, and J.~Henderson, ``The pressure tensor
  at the planar surface of a liquid,'' {\em Molecular physics}, vol.~48, no.~6,
  pp.~1357--1368, 1983.

\bibitem{tolman1949effect}
R.~C. Tolman, ``The effect of droplet size on surface tension,'' {\em The
  journal of chemical physics}, vol.~17, no.~3, pp.~333--337, 1949.

\bibitem{rowlinson2013molecular}
J.~S. Rowlinson and B.~Widom, {\em Molecular theory of capillarity}.
\newblock Courier Corporation, 2013.

\bibitem{montero2019interfacial}
P.~Montero~de Hijes, J.~R. Espinosa, E.~Sanz, and C.~Vega, ``Interfacial free
  energy of a liquid-solid interface: Its change with curvature,'' {\em The
  Journal of Chemical Physics}, vol.~151, no.~14, p.~144501, 2019.

\bibitem{espinosaJCP2016_2}
J.~R. Espinosa, J.~M. Young, H.~Jiang, D.~Gupta, C.~Vega, E.~Sanz, P.~G.
  Debenedetti, and A.~Z. Panagiotopoulos, ``On the calculation of solubilities
  via direct coexistence simulations: Investigation of nacl aqueous solutions
  and lennard-jones binary mixtures,'' {\em The Journal of Chemical Physics},
  vol.~145, no.~15, p.~154111, 2016.

\bibitem{PRL_2005_94_018104}
R.~J. Allen, P.~B. Warren, and P.~R. ten Wolde, ``Sampling rare switching
  events in biochemical networks,'' {\em Phys. Rev. Lett.}, vol.~94, p.~018104,
  2005.

\bibitem{JCP_1992_96_04655}
J.~S. van Duijneveldt and D.~Frenkel, ``Computer simulation study of free
  energy barriers in crystal nucleation,'' {\em J. Chem. Phys.}, vol.~96,
  p.~4655, 1992.

\bibitem{filion:244115}
L.~Filion, M.~Hermes, R.~Ni, and M.~Dijkstra, ``Crystal nucleation of hard
  spheres using molecular dynamics, umbrella sampling, and forward flux
  sampling: A comparison of simulation techniques,'' {\em J. Chem. Phys.},
  vol.~133, no.~24, p.~244115, 2010.

\end{thebibliography}

\end{document}